\documentclass[11pt]{article}\topmargin=-0.7cm \textheight 222mm \textwidth=16.8cm\oddsidemargin=2mm %\date{\  }% for xxx
\newcommand{\comm}[1]{}

\usepackage{amssymb}
\usepackage{amsmath}\usepackage{graphicx} \usepackage[numbers]{natbib}\usepackage{epsfig}
\def\citet{\cite}

\def\xxxonly{\comm}
\def\xxxonly{ }
\def\noxxx{\comm}
\usepackage{times}\hfuzz=10pt 

 \usepackage{color}
\newtheorem{theorem}{Theorem}[section]

\newtheorem{proposition}[theorem]{Proposition}
\newtheorem{lemma}[theorem]{Lemma}
\newtheorem{corollary}[theorem]{Corollary}

\newtheorem{definition}[theorem]{Definition}
\newtheorem{remark}[theorem]{Remark}
\newtheorem{example}[theorem]{Example}
\def\e{\varepsilon}

\def\defi{\stackrel{{\scriptscriptstyle \Delta}}{=}}

\def\OO{{\scriptscriptstyle O}}

\def\a{\alpha}
\def\d{\delta}
\def\o{\omega}

\def\Y{{\cal Y}}

\def\w{\widehat}
\def\Ind{{\mathbb{I}}}

\def\mes{{\rm mes\,\!}}

\def\Re{{\rm Re\,\!}}

\def\R{{\bf R}}

\def\Z{{\cal Z}}
\def\ZZ{{\bf Z}}
\def\PP{{\cal P}}

\def\b{\beta}
\def\s{\delta}
\def\g{\gamma}
\def\C{{\bf C}}
\def\W{{\cal W}^*}
\def\W{{\cal W}}
\def\ww{\widetilde}
\def\X{{\cal X}}
\def\t{\theta}
\def\oo{\bar}
\def\s{\sigma}
\def\V{{\cal V}}

\def\M{{\cal M}}

\def\T{{\mathbb{T}}}

\newcommand{\be}{\begin{equation}}
\newcommand{\ee}{\end{equation}}
\newcommand{\bd}{\begin{displaymath}}
\newcommand{\ed}{\end{displaymath}}
\newcommand{\ba}{\begin{array}{ll}}
\newcommand{\ea}{\end{array}}
\newcommand{\baa}{\begin{eqnarray}}
\newcommand{\eaa}{\end{eqnarray}}
\newcommand{\baaa}{\begin{eqnarray*}}
\newcommand{\eaaa}{\end{eqnarray*}}
\font\sm=cmr10
\def\PP{{\cal P}}

\def\oo{\bar}

\def\a{\alpha}

\def\K{{\cal K}}
\def\yo{y}%{K_0}
\def\ew{\left(e^{i\o}\right)}

\def\T{{\mathbb{T}}}
\def\ZZ{{\mathbb{Z}}}

\def\TT{{\cal T}}
\def\HHH{{\rm H}}

\def\ee{\epsilon}
\def\V{{\cal V}}

\def\dm{{m\nu}}
\def\SS{{decim}\,}
\def\SSS{{ subseq}\!}
\def\SupS{{superseq}\,}

\def\qq{q}
\def\WW{\varrho}
\date{Submitted: March 6 2018. Revised: August 19 2018}%xxx
\title{Predictability of sequences and subsequences with spectrum degeneracy at periodically located points}
\author{
Nikolai Dokuchaev}
\begin{document}
 \vspace{-0.5cm}
\def\brea{}
\def\breakk{}
\def\break{}
\maketitle %\rd{Better proof. Nemnogo bol'she chem v braided.}

\begin{abstract}
The paper established sufficient conditions of predictability
with degeneracy for the spectrum at $M$-periodically located isolated points on the unit circle.
It is also shown that  $m$-periodic subsequences of these sequences are also predictable
if $m$ is a divisor of $M$. \index{In addition, it is shown that compound sequences formed from these
subsequences are also predictable.} The predictability can be achieved for finite horizon with linear predictors  defined by convolutions with certain kernels. As an example of applications, it is shown that there exists a class of sequences that is everywhere dense
in the class of all square-summable sequences and such that its members can be recovered from their periodic
subsequences. This recoverability is associated with certain spectrum degeneracy of a new kind.

% 42A38, %       Fourier and Fourier-Stieltjes transforms and other transforms of Fourier type
%562M15,      % Spectral analysis is stochastic processes
%42B30  %Hardy spaces
\end{abstract}
\section{Introduction}
It is well known that band-limited sequences, or sequences  with Z-transform vanishing on an open arc of the unit circle
$\T=\{z\in\C:\ |z|=1\}$, are predictable, i.e. their future values can be estimated with arbitrarily small
error from observation of their history; see e.g. the literature review in \cite{D12a}.
It is also known that
\begin{itemize}
\item A sequence is predictable if its Z-transform vanishes with  a certain rate at a single point \cite{D12a}.
\item If there are $m$-periodic spectrum gaps  on $\T$ of positive measure, then  $m$-periodic subsequences are predictable.
\end{itemize}

The paper established sufficient conditions of predictability
with degeneracy for the spectrum at $M$-periodically located isolated points on the unit circle
(Theorem \ref{ThV}).
In addition, it  shows that $m$-periodic subsequences of these sequences are also predictable
if $m$ is a divisor of $M$ (Corollary \ref{corrM}). This is a new result for the case
where either  $M>m$ or the spectrum does not have gaps of positive measure.

It is shown that the corresponding  predictability can be achieved for finite horizon with linear predictors  defined by convolutions with certain kernels that were presented  explicitly.

This result implies sequences  are predictable if they are formed as a combination of sequences with  periodically located isolated points  on $\T$ where  Z-transform is vanishing (Theorem \ref{ThM2}). These compound  sequences feature predictability even if
their Z-transform is separated from zero on $\T$.

 As an example of applications, it is shown that there exists a class of sequences that is everywhere dense
in the class of all square-summable sequences and such that its members can be recovered from their periodic
subsequences (Theorems \ref{ThDeg}-\ref{ThDense}). These theorems represents a modification of the related result obtained in \cite{D16x}.

\subsubsection*{Some definitions and notations}
We denote by $L_2(D)$ the usual Hilbert space of complex valued
square integrable functions $x:D\to\C$, where $D$ is a domain.

We denote by $\ZZ$  the set of all integers. Let $\ZZ_q^+=\{k\in\ZZ:\ k\ge q\}$, let $\ZZ_q^-=\{k\in\ZZ:\ k\le q\}$, and let $\ZZ_{[a,b]}=\{k\in\ZZ:\ a\le a\le b\}$.

\par
For a set $G\subset \ZZ$ and $r\in[1,\infty]$,
we denote by $\ell_r(G)$ a Banach
space of complex valued sequences $\{x(t)\}_{t\in G}$
such that
$\|x\|_{\ell_r(G)}\defi \left(\sum_{t\in G}|x(t)|^r\right)^{1/r}<+\infty$ for $r\in[1,+\infty)$,
and $\|x\|_{q(G)}\defi \sup_{t\in G}|x(t)|<+\infty$ for $r=\infty$.
We denote  $\ell_r=\ell_r(\ZZ)$.

We denote by $B_r(\ell_2)$ the closed ball of radius $r>0$ in $\ell_2$.

\par
Let $D^c\defi\{z\in\C: |z|> 1\}$, and let $\T\defi \{z\in\C: |z|=1\}$.
\par
For  $x\in \ell_2$, we denote by $X=\Z x$ the
Z-transform  \baaa X(z)=\sum_{k=-\infty}^{\infty}x(k)z^{-k},\quad
z\in\C. \eaaa Respectively, the inverse Z-transform  $x=\Z^{-1}X$ is
defined as \baaa x(k)=\frac{1}{2\pi}\int_{-\pi}^\pi
X\left(e^{i\o}\right) e^{i\o k}d\o, \quad k=0,\pm 1,\pm 2,....\eaaa
For $x\in \ell_2$, the trace $X|_\T$ is defined as an element of
$L_2(\T)$.

 Let  $\HHH^r(D^c)$ be the Hardy space of functions that are holomorphic on
$D^c$ including the point at infinity  with finite norm
\baaa
\|h\|_{\HHH^r(D^c)}=\sup_{\rho>1}\|h(\rho e^{i\o})\|_{L_r(-\pi,\pi)},
\eaaa
 $r\in[1,+\infty]$.

\index{Note that Z-transform defines a bijection
between the sequences from $\ell_2^+$ and the restrictions (i.e.,
traces) $X|_{\T}$ of the functions from $\HHH^2(D^c)$ such that  $\overline{X\ew} =X\left(e^{-i\o}\right)$ for $\o\in\R$; see, e.g., \cite{Linq}, Section 4.3.
If $X\ew\in L_1(-\pi,\pi)$ and $\overline{X\ew} =X\left(e^{-i\o}\right)$, then $x=\Z^{-1}X$
is defined as an element of $\ell_\infty^+$.}

For $n\in \ZZ_1^+$ and $\b\in(-\pi,\pi]$, let $R_{n,\b}$ be the set of all $\o\in (-\pi,\pi]$ such that $e^{i\o n}=e^{i\b}$, i.e.
\baaa
R_{n,\b}=\{\o_{n,\b,k}\}_{k=0,...,n-1},\quad \hbox{where}\quad   \o_{n,\b,k}=\frac{2\pi k-\b}{ n}.
\eaaa

In this papers, we focus on processes without spectrum gaps
of positive measure on $\T$ but with spectrum gaps at isolated periodic points on $\T$ where Z-transforms vanishes with a certain rate.

Let $L>1$ be given. (In fact, we can take $L=1$ for all results below except  Theorem \ref{ThDeg}, where we require that $L>1$  is sufficiently large;
this is rather technical conditions that is used in the proof.)
\par
For  $\b\in (-\pi,\pi]$,  $q>1$, $c>0$, and $\o\in(-\pi,\pi]$, set \baaa
\WW(\o,\ww\o,q,c)=\frac{1}{L}\max\left(L,\exp\frac{c}{|e^{i\o}-e^{i\ww\o}|^{q}}\right).
\label{hdef}\eaaa
\comm{
For $q>1$, $c>0$, and $\o\in(-\pi,\pi]$, set \baaa
\WW(\o,q,c)=\exp\frac{c}{[(\cos(\o)+1)^2+\sin^2(\o)]^{q/2}}.
\label{hdef}\eaaa}

%We denote by $B_\rho(\ell_2)$ the closed ball of radius $\rho>0$ in $\ell_2$.

For $r>0$, let $\X_{n,\b}(q,c,r)$ be the set of all   $x\in \ell_2$ such that
\baa \max_{k=0,1,...,n-1}\int_{-\pi}^\pi |X\ew|^2 \WW(\o,\o_{n,\b,k},q,c)^2d\o\le r,
\label{hfin}\eaa\ where $X=\Z x$.

  Let
\baaa
\X_{n}\defi \cup_{\b\in(-\pi,\pi],q>0,c>0,r>0} \X_{n,\b}(q,c,r).
\eaaa

For $m\in \ZZ_1^+$ and $s\in\ZZ$, let $\SS_{m,s}:\ell_2\to\ell_2$ be a mapping representing
decimation of a sequence such that
 $y(t)=x(t)\Ind_{\{(t+s)/m\in\ZZ\}}$ for $y=\SS_{m,s}x$.

 Let $\SSS_{m,s}:\ell_2\to\ell_2$ be a mapping representing
extraction of a subsequence such that $\w y(k)=x(km-s)$ for all $k\in\ZZ$ for $\w y=\SSS_{m,s}x$.

Let $\SupS_{m,s}:\ell_2\to\ell_2$ be a mapping representing
construction of a supersequence such that
 $\oo x(t)=\oo y(t)\Ind_{\{(t+s)/m\in\ZZ\}}$ for $\oo x=\SupS_{m,s}\oo y$.

\def\wh{h}\def\wH{H}
%\begin{definition} %Let either $r=2$ or $r=+\infty$.
Let $\w\K^+$ be the class of functions $\wh:\ZZ\to\R$
such that $\wh (t)=0$ for $t<0$ and such that $\wH=\Z\wh\in H^\infty(D^c)$. Let $\w\K^-$ be the class of functions $\wh:\ZZ\to\R$
such that $\wh (t)=0$ for $t>0$ and such that $\wH=\Z\wh\in H^\infty(D)$. Let $\w\K$ be the linear span of $\w\K^+\cup\w\K^-$.
%\end{definition}

\section{The main results}
\label{SecM}
\def\RR{{\cal R}}
Up to the end of this paper, we assume that we are given $m\ge 1$, $m\in\ZZ$.

\begin{definition}\label{defP}
 Let $\X\subset \ell_2$ be a class of
processes.
 We say that the class $\X$ is  uniformly $\ell_2$-predictable on finite horizon   if for any $\e>0$ and any integers $n>0$ (any $n <0$) there exists $\wh\in \w\K^{+}$ (respectively, $\wh\in \w\K^{-}$ for $n<0$) and $\psi\in\ell_\infty$  such that $\inf_k|\psi(k)|>0$, $\sup_k|\psi(k)|<+\infty$, and   \baaa \sum_{t\in\ZZ}|x(t+n)- \w x(t)|^2\le \e\quad
\forall x\in\X, \label{predu}\eaaa where \baaa
\w x(t)\defi
\psi(t)^{-1}\sum_{s\in\ZZ}\wh(t-s)\psi(s)x(s).
\eaaa
\end{definition}

In Definition \ref{defP}, the use of $\wh\in\w \K^-$ means causal extrapolation, i.e. estimation  of $\w x(t+n)$
for $n>0$ is based on observations $\{x(k)\}_{k\le t}$, and the use of $\wh\in\w \K^+$ means anti-causal extrapolation, i.e. estimation  of $\w x(t+n)$ for $n<0$
is based on observations  $\{x(k)\}_{k\ge t}$.

\begin{theorem}\label{ThV}
For any $\qq>0$, $r>0$, $\nu\in \ZZ_1^+$, $\b\in(-\pi,\pi]$, $s\in\ZZ$,  and $\rho>0$, the class of sequences
$\X_{m\nu,\b}(\qq,c,r)$ is uniformly predictable on finite horizon in the sense of Definition
\ref{defP}  with $\psi(t)=e^{i\t t}$ and $\t=(\b-\pi)/(m\nu)$.
\end{theorem}
\subsection{Robustness of prediction}
\begin{definition}\label{defRob}  Let $\X\subset \ell_2$ be a set of sequences.
Consider  a problem of predicting  $\{x(k)\}_{\in\ZZ_{[-M,M]}}$  using observations of noise contaminated sequences $x=\ww x+\xi$, where  $\ZZ_{[-M,M]}=\{k\in\ZZ,\ |k|\le M\}$,
 $\ww x\in\X$, and where $\xi\in\ell_2$ represents a noise. Suppose that only
truncated traces of observations of
$\{x(k),\ k\in\TT_0,\ -N\le k<-M\}$ available (or only  traces $\{x(k),\ k\in\TT_0,\ -N\le k<-M\}$ are available), where  $N>0$  is an integer. We say that the class $\X$ allows uniform and robust prediction if,  for   any integer $M>0$ and any $\e>0$,  there exists $\rho >0$,  $N_0>0$,
a  set of sequences $\{\psi_{t}(\cdot)\}_{t\in\ZZ_{[-M,M]}}$ such that $\psi_t\in\ell_\infty$, $\inf_k|\psi_t(k)|>0$, $\sup_k|\psi_t(k)|<+\infty$ for any $t$, and a set
 $\{\wh_{t}(\cdot)\}_{t\in\ZZ_{[-M,M]}}\subset \K^+$ (or a set
 $\{\wh_{t}(\cdot)\}_{t\in\ZZ_{[-M,M]}}\subset \K^-$ respectively) such that \baaa
\max_{t\in\ZZ_{[-M,M]}}| x(t)-\w x(t)|\le \e \quad
\forall \ww x\in\X_{m\nu,\b}(\qq,c,r),\quad\forall  \xi\in B_\rho(\ell_2), \label{predUU}\eaaa
  for any  $N>N_0$ and for  \baaa
\w x(t)=
\psi_t(t)^{-1}\sum_{s\in \TT_0,\ M<|s|\le N}\wh_{t}(t-s)\psi_t(s)x(s),\quad t\in\ZZ_{[-M,M]},\eaaa
 with $\psi_t(s)=e^{i\t s}$ for all $t$ and $\t=(\b-\pi)/(m\nu)$.
  \end{definition}

The following theorem shows that predicting  is robust with respect to noise contamination and truncation.
\begin{theorem}\label{ThR} For given $\nu\in\ZZ_1^+,\b\in(-\pi,\pi],\qq>0,c>0,r>0$, the class of sequences  $\X_{m\nu,\b}(\qq,c,r)$ allows uniform and robust prediction  in the sense of Definition \ref{defRob}  with $\psi(t)=e^{i\t t}$ and $\t=(\b-\pi)/(m\nu)$.   \end{theorem}

\subsection{Predictability of subsequences}

\begin{definition}\label{defADN}  Let $\qq>0$,  $\nu\in\ZZ_1^+$, $\b\in(-\pi,\pi]$, $q>0$, $c>0$, $r\ge 0$, and  $s\in\ZZ$,
 be given.
 \begin{enumerate}
 \item
 Let $\Y_{m,\nu,\b,s}(\qq,c,r)$ be the set of all $y\in \ell_2$
 such that there exists   $x\in \X_{m\nu,\b}(\qq,c,r)$ such that  $y=\SS_{m,s} x$.
 \item
Let $\w\Y_{m,\nu,\b,s}(\qq,c,r)$ be the set of all $y\in\ell_2$
such that there exists   $x\in \X_{m\nu,\b}(\qq,c,r)$ such that
$\w y=\SSS_{m,s}x$ for $x\in \X_{m\nu,\b}(\qq,c,r)$.
 \end{enumerate}
  \end{definition}
\begin{lemma}\label{lemmaSS} $\Y_{m,\nu,\b,s}(\qq,c,r)\subset\X_{m\nu,\b}(\qq,c,r_1)$ for some $r_1=r_1(m,\nu,\b,\qq,s,r)$.
\end{lemma}
\begin{corollary}\label{corrM} Let  $\nu\in\ZZ_1^+$, $\b\in(-\pi,\pi]$, $q>0$, $c>0$, $r\ge 0$, and  $s\in\ZZ$,
 be given.
Let either $\X=\Y_{m,\nu,\b,s}(\qq,c,r)$ or  $\X=\w\Y_{m,\nu,\b,s}(\qq,c,r)$. Then  the class of sequences
$\X$ is uniformly predictable on finite horizon in the sense of Definition
\ref{defP} and allows robust uniform prediction in the sense of Definition
\ref{defRob}. This predictability is robust with respect to noise contamination in the sense
of Theorem \ref{ThR}.
\end{corollary}

\par

  Let
\baaa
\Y_{m,s}\defi \cup_{\b\in(-\pi,\pi],\nu\in\ZZ_1^+,\qq>0,c>0,r>0} \Y_{m,\nu,\b,s}(\qq,c,r),\quad
\Y_m\defi \cup_{s\in\ZZ} \Y_{m,s}
\eaaa
and
\baaa
\w\Y_{m,s}\defi \cup_{\b\in(-\pi,\pi],\nu\in\ZZ_1^+,\qq>0,c>0,r>0}\w \Y_{m,\nu,\b,s}(\qq,c,r),\quad
\w\Y_m\defi \cup_{s\in\ZZ} \w\Y_{m,s}.
\eaaa
\begin{corollary}\label{corrSUB} Let either $\Y=\Y_{m}$ or  $\Y=\w\Y_{m}$. For any $p\in\ZZ$, any sequence   $y\in\X$ is uniquely defined by its tail  $\{y(k)\}_{k\in  \ZZ^-_p}$. It is also uniquely defined
  by its tail  $\{y(k)\}_{k\in\ZZ_p^+}$.
\end{corollary}
\begin{corollary}\label{corr2}
 $\ell_2\setminus \w\Y_m \neq \emptyset$.
\end{corollary}

Theorem \ref{ThV} and Corollary \ref{corrSUB}  indicate that  sequences from $\Y_{m}$ and $\w\Y_{m}$  feature predictability that is usually associated with spectrum degeneracy. By Lemma \ref{lemmaSS},
it is expectable for $\Y_m$ since Z-transforms of sequences from  $\Y_{m}$
vanish at a periodic set of point of $\T$ as was defined for sequences from $\X_{m\nu,\b}(\qq,c,r)$. However,
it is less expectable for sequences from  $\w\Y_{m}$ since their Z-transforms can be actually separated from zero on $\T$.
\subsection{On detecting predictable sequences}
Sequences from  $\w\Y_{m}$ are predictable;  predictability of these processes  is defined by certain  spectrum degeneracy of their supersequences.  By Corollary \ref{corr2}, there are sequences in $\ell_2$ that do not belong to this class.

For a given $\w y\in\ell_2$, the question arises if $\w y\in \w\Y_m$. The following lemma suggests an answer for this question.

\begin{lemma}\label{lemmaSD}  For $\w y\in\ell_2$ and $s\in\ZZ$, we  have that  $\w y\in\w\Y_{m,s}$  if and only if
 $y= \SupS_{m\nu,s}\w y\in\X_{m\nu}$ for some $\nu\in\ZZ_1^+$.
 \end{lemma}
According to this Lemma, $\w y$ is a subsequence of a sequence $x$ with degenerate spectrum if and only if the supersequence $y$ obtained from $x$ by
nullifying members not belonging to $\w y$ also has a degenerate spectrum.
\subsection{Predictability of compound processes}

For $m\in\ZZ_1^+$, $\qq>0$, $c>0$,  $\nu=(\nu_0,...,\nu_{m-1})\in(\ZZ_1^+)^m$, $\b=(\b_0,...,\b_{m-1})\in(-\pi,\pi]^m$, and  $s\in\ZZ$, let
$\Y^{comp}_{m,\nu,\b,s}(\qq,c,r)$ be the set of all sequences $y$ such that
\baaa
y(k)=\sum_{d=0}^{m-1}\xi_d\left(k+d\right)\Ind_{\left\{ \frac{k+d}{m}\in\ZZ\right\}},
\label{xPmC}
\eaaa
where $\xi_d\in\Y_{m,\nu_d,\b_d,s}(\qq,c,r)$.

\begin{theorem}\label{ThM2} Let $m\in\ZZ_1^+$, $\qq>0$,  $\nu\in(\ZZ_1^+)^m$, $\b\in(-\pi,\pi]^m$, $\qq>0$, $c>0$, and  $s\in\ZZ$,
 be given. Then  the class of sequences $\X=\Y^{comp}_{m,\nu,\b,s}(\qq,c,r)$
 is uniformly predictable on finite horizon in the sense of Definition
\ref{defP} with $\psi_t(s)=e^{i\t_t s}$ and $\t_t=(\b_d-\pi)/(m\nu_d)$ if $(t+d)/m\in\ZZ$.
\end{theorem}

  Let
\baaa
&&\Y^{comp}_{m,s}\defi \cup_{\b\in(-\pi,\pi],\nu\in\ZZ_1^+,\qq>0,c>0,r>0} \Y^{comp}_{m,\nu,\b,s}(\qq,c,r),\quad
\breakk \Y^{comp}_m\defi \cup_{s\in\ZZ} \Y^{comp}_{m,s}.
\eaaa
\begin{corollary}\label{corrSUB2}  For any $p\in\ZZ$, any sequence   $\Y^{comp}_m$ is uniquely defined by its tail  $\{y(k)\}_{k\in  \ZZ^-_p}$. It is also uniquely defined
  by its tail  $\{y(k)\}_{k\in\ZZ_p^+}$.
\end{corollary}
\begin{corollary}\label{corr22}
 $\ell_2\setminus \Y _m^{comp} \neq \emptyset$.
\end{corollary}
It can be noted that Z-transform of sequences from $\Y _m^{comp}$ can be separated from zero on $\T$, so their spectrum is non-degenerate in the usual sense.
\section{On recovery of sequences from their subsequences}
We will need some modification of Definition \ref{defR}.
\begin{definition}\label{defR}
Let $\X\subset \ell_2$ be a class of sequences. Let $\ww\TT\subset \ZZ$.
We say that the class $\X$ is  uniformly recoverable  from observations of $x$ on  $\ww\TT$  if, for any integer $M>0$ and any $\e>0$, there exists
a  set of sequences $\{\psi_{t}(\cdot)\}_{t\in\ZZ_{[-M,M]}}$ such that $\psi_t\in\ell_\infty$, $\inf_k|\psi_t(k)|>0$, $\sup_k|\psi(k)|<+\infty$ for any $t$, and
a set $\{\wh_{t}(\cdot)\}_{t\in\ZZ_{[-M,M]}} \subset \w\K$ such that \baaa
\max_{t\in\ZZ_{[-M,M]}} |x(t)- \w x(t)|\le \e\quad
\forall x\in\X, \label{predU}\eaaa where  \baaa
\w x(t)=\psi_t(t)
\sum_{s\in \ww\TT}\wh_{t}(t-s)\psi_t(s)x(s),\quad t\in\ZZ_{[-M,M]}.\eaaa
\end{definition}
In Definition \ref{defR}, the use of $\wh\in\w \K^-$ means causal extrapolation, i.e. selection of $\w x(t+n)$
for $n>0$ based on observations $\{x(k)\}_{k\le t}$, and the use of $\wh\in\w \K^+$ means anti-causal extrapolation, i.e. selection of $\w x(t+n)$ for $n<0$
based on observations  $\{x(k)\}_{k\ge t}$.

\subsection*{A special type of spectrum degeneracy}

\begin{definition}\label{defDeg}
Let $\qq>0$, $r>0$, $\b\in(-\pi,\pi]$, $s\in\ZZ$,  and $\rho>0$ be given.
 We say that $x\in \ell_2$
 features braided \index{tangled} spectrum degeneracy with parameters  $m,r,\b,s,\qq$   if, for $d=-m+1,...,m-1$,
 there exist $\nu=(\nu_{-m+1},...,\nu_{m-1})\in(\ZZ_1^+)^{2m-1}$ and  $\xi_d\in
 \X_{m\nu_d,\b}(\qq,c,r)$
 \index{\W_{m,\nu_d,\pi,s}(\d,r)}
 such that the following  holds:
 \begin{enumerate}
\item The sets $R_{m\nu_d,\b}$ are mutually disjoint for $d=-m+1,....,m-1$.
\item
\baa
 &&\xi_d(k)=\xi_0(k),\quad k\le 0,\quad d>0\nonumber\\
 && \xi_d(k)=\xi_0(k),\quad k\ge 0,\quad d<0,
\label{yPm} \eaa
\item
 \baa
 &&x(k)=\sum_{d=0}^{m-1}\xi_d\left(k+d\right)\Ind_{\left\{ \frac{k+d}{m}\in\ZZ\right\}},\quad k\ge 0,\nonumber\\
 &&x(k)=\sum_{d=-m+1}^{0}\xi_d\left(k+d\right)\Ind_{\left\{ \frac{k+d}{m}\in\ZZ\right\}},\quad k\le 0.
\label{xPm} \eaa
\end{enumerate}
We denote by $\PP_{m,\nu,\b}(\qq,c,r)$ the set  of  all  sequences  $x$ with this feature, and we denote  $\PP_m=\cup_{m,\nu,\b,r,\qq,c,r}\PP_m(\qq,c,r)$.
  \end{definition}
\begin{example}\label{ex1} A possible choice of integers $\nu_d$  such that all sets $R_{m\nu_d,\pi}$ are mutually disjoint is  $\nu_d=2^d$ for $d\ge 0$ and   $\nu_d=2^{2m+d-1} $ for $d<0$.
 \end{example}
 Let
\baaa
& \TT_0=\{k\in \ZZ:\ k/m\in\ZZ\},\quad &\TT_0(s)=\{k\in \TT:\ |k|>s\},\\
 &\TT_d=\{k\in \ZZ:\ (k-d)/m\in\ZZ,\ k>0\},\quad &d=1,...,m-1,\\
  &\TT_d=\{k\in \ZZ:\ (k-d)/m\in\ZZ,\ k<0\},\quad &d=-m+1,...,-1.
 \eaaa
\begin{theorem}\label{ThDeg} Let $\d>0$ and  $s>0$ be given, and $L>1$ be sufficiently large in the definition for $\WW$. Then the class  $\PP_{m,\nu,\b}(\qq,c,r)$ is uniformly recoverable from observations on $\TT_0(s)$
in the sense of Definition \ref{defR}.
\end{theorem}

\begin{remark} {\rm As is seen from the proof below, Theorem \ref{ThDeg} is still valid if $\w\K$
in  Definition \ref{defR} is replaced by a smaller set $\w\K^-\cup \w\K^+$.}
\end{remark}
\begin{corollary}\label{corrU}
 A sequence   $x\in\PP_m$ is uniquely defined by its subsequence $\{x(km)\}_{k\in \ZZ}$.  Moreover, for any $s>0$, a sequence   $x\in\PP_m$ is uniquely defined by its subsequence $\{x(km)\}_{k\in \ZZ,\ |k|>s}$.
\end{corollary}

%\subsubsection*{The set $\PP_m$ is everywhere dense in $\ell_2$}

\begin{theorem}\label{ThDense}
For any $x\in \ell_2$ and any $\e>0$, there exists  $\w x\in \PP_{m,\nu,\b}(\qq,c,r)$  such that \baa
\|x-\w x\|_{\ell_2}\le \e.
\label{xdd}\eaa
In other words, the  set $\PP_m$ is everywhere dense in $\ell_2$.
\end{theorem}

Let us compare these results with the result of \cite{Ser}, where a method was suggested
 for recovery from observations of subsequences
of missing values  oversampling sequences
for band-limited continuous time  functions.  The  method \cite{Ser} is also applicable
for general type band-limited sequences from $\ell_2$.
The algorithm in  \cite{Ser} requires to observe quite large number of subsequences, and this number is increasing as the size of spectrum gap  on $\T$ is decreasing.
Theorems \ref{ThDeg} and \ref{ThDense} ensures recoverability with   just one subsequence for a class of sequences that everywhere dense in $\ell_2$.

\subsubsection*{Robustness of recovery}

The following theorem shows that recovery of a finite part of a sequence from $\PP_m$
from its $m$-periodic subsequence is robust with respect to noise contamination and truncation.

\begin{theorem}\label{ThRR} Let $m,\nu,\b,r,\qq,c,r$  be given.
Consider  a problem of recovery of  the set  $\{x(k)\}_{k=-M}^M$   from observed subsequences of
 noise contaminated sequences $x=\ww x+\xi$, where  $M\in\ZZ_0^+$,  $\ww x\in\PP_{m,\nu,\b}(\qq,c,r)$ and where $\xi\in\ell_2$ represents a noise. Suppose that only
truncated traces of observations of
$\{x(k),\ k\in\TT_0,\ M<|k|\le N\}$ are available, where  $N>0$  is an integer.
Then for   any integer $M>0$ and any $\e>0$,  there exists $\rho >0$,  $N_0>0$,
a  set of sequences $\{\psi_{t}(\cdot)\}_{t\in\ZZ_{[-M,M]}}$ such that $\psi_t\in\ell_\infty$, $\inf_k|\psi_t(k)|>0$, $\sup_k|\psi(k)|<+\infty$ for any $t$, and a set
 $\{\wh_{t}(\cdot)\}_{t\in\ZZ_{[-M,M]}}\subset \w\K$ such that \baaa
\max_{t\in\ZZ_{[-M,M]}}| x(t)-\w x(t)|\le \e \quad
\forall \ww x\in\PP_{m,\nu,\b}(\qq,c,r),\quad\forall  \xi\in B_\rho(\ell_2), \label{predUU2}\eaaa
  for any  $N>N_0$ and for  \baaa
\w x(t)=\psi_t(t)
\sum_{s\in \TT_0,\ M<|s|\le N}\wh_{t}(t-s)\psi_t(s)x(s),\quad t\in\ZZ_{[-M,M]}.\eaaa
 \end{theorem}

\section{Proofs} It suffices to consider prediction on $n$ steps forward
with $n>0$. The case where $n<0$ can be considered similarly.

\subsubsection*{Special predicting kernels}

The proofs are based on special predicting kernels  and their transfer functions representing modification
of kernels and transfer functions  introduced in  \cite{D12a}.

Let us introduce transfer functions and its inverse Z-transform
   \baa
 \w H_n(z)=\w H_{n,\dm}(z) \defi z^n V(z^{\dm })^n,\quad z\in\C,\qquad \w h_n=\Z^{-1}\w H_n, \label{wH}\eaa
where
 \baaa
 V(z)\defi 1-\exp\left[-\frac{\g}{z+ 1-\g^{-\w r}}\right],
\label{wK}
\eaaa
and where $\w r>0$ and $\g>0$ are  parameters. \index{Let $r > 2/(q-1)$ for some $q>1$.}
 Functions $V$ were introduced in  \cite{D12a}.
 In the notations from \cite{D12a}, $\w r=2\mu/(q-1)$, where $\mu>1$ and $q>1$ are parameters; $q$ describes  the required  rate of spectrum degeneracy.
We assume that $r$ is fixed, and we consider variable $\g\to +\infty$.

In the proof below, we will show that $\w H_n\ew$ approximates $e^{in\o}$ and therefore defines
a linear $n$-step predictor with the kernel $\w h_n=\Z^{-1}\w H_n$.

\subsubsection*{The case where $\b=\pi$ }
\begin{lemma}\label{lemmaX} Theorem \ref{ThV} holds for the case where $\b=\pi$. \end{lemma}

{\em Proof of Lemma \ref{lemmaX}}.
The proof represents a modification of the proof from
\cite{D12a}, where the case of $n=1$ was considered and where
the spectrum degeneracy were assumed to be at a single point only. In addition,  represents a modification of the proof from \cite{D16x}, where the case of $\b=\pi$ was considered.

\def\OO{W}
Let $\a=\a(\g)\defi 1-\g^{-r}$. Clearly, $\a=\a(\g)\to 1$  as $\g\to+\infty$.

Let $\OO(\a)=\arccos(-\a)$, let $D_+(\a)=(-\OO(\a),\OO(\a))$, and let
 $D(\a)\defi[-\pi,\pi]\backslash D_+(\a)$. We have that  $\cos(\OO(\a))+\a=0$, $\cos(\o)+\a>0$ for  $\o\in D_+(\a)$,
and $\cos(\o)+\a<0$ for  $\o\in D(\a)$.
\par

It was shown in \citet{D12a} that the following holds:
\begin{itemize}
\item[(a1)] $V(z)\in \HHH^{\infty}(D^c)$ and $zV(z)\in
\HHH^{\infty}(D^c)$.
\item[(a2)] $V(e^{i\o})\to 1$ for all  $\o\in (-\pi,\pi)$ as  $\g\to +\infty$.
\item[(a3)] If $\o\in D_+(\a)$  then $\Re \left(\frac{\g}{e^{i\o}+\a}\right) >0$ and $|V\ew -1|\le 2$.
    \item[(a4)] For any $c>0$,  there exists $\g_0>0$
    such that, for any $\g\ge \g_0$,
\baaa
|V\ew-1|
\WW(\o,\pi,q,c)^{-1}\le 1 \quad  \forall\o\in D(\a).
\eaaa
\end{itemize}

Without a loss of generality, we assume below that $\g>\g_0$ for corresponding $c$.

Let
\baaa
&&Q(\a)=\cup_{k=0}^{\dm-1} \left(\frac{\OO(\a)+2\pi k}{\dm},\frac{2\pi-\OO(\a)+2\pi k}{\dm} \right),\qquad \breakk
Q_+(\a)=[-\pi,\pi]\setminus Q(\a).\eaaa
\index{For example, if $m=2$ then
\baaa
&&Q(\a)=(\OO(\a)/2,\pi-\OO(\a)/2)\cup (-\OO(\a)/2,\OO(\a)/2-\pi),\qquad \breakk
Q_+(\a)=[-\pi,\pi]\setminus Q(\a).\eaaa}

From the properties of $V$, it follows that the following holds.
\begin{itemize}
\item[(b1)] $V(z^{\dm})^n\in \HHH^{\infty}(D^c)$ and $ \w H_n(z)= z^nV(z^{\dm})^n\in
\HHH^{\infty}(D^c)$.
\item[(b2)] $V\left(e^{i \o \dm }\right)\to 1$  and $\w H_n\ew\to e^{in\o}$ for all  $\o\in (-\pi,\pi]\setminus R_{\dm}$ as  $\g\to +\infty$.
\item[(b3)]
\baaa
\Re \left(\frac{\g}{e^{i \o \dm }+\a}\right) >0,\quad
\left|V\left(e^{i\o \dm }\right) -1\right|\le 1 \quad\brea \forall \o\in Q_+(\a).
\eaaa
   \item[(b4)]
   For any $c>0$, there exists $\g_0>0$
    such that, for any $\g>\g_0$,
\baaa
|V\left(e^{i\o \dm }\right)-1|
\WW(\o,\o_{\dm,\pi,k},q,c)^{-1}d\o\le 1\quad \forall\o\in Q(\a), \forall k=0,1,...,\dm-1.
\eaaa
\end{itemize}
\par
Let $p=\{p_l\}\in \R^n$ be such that $a^n-1=\sum_{l=0}^n p_l(a-1)^l$. In this case,
\baaa
V\left(e^{i\o \dm }\right)^n-1= \sum_{l=0}^n p_l (V\left(e^{i\o \dm }\right)-1)^l
\eaaa
and
\baaa
|V\left(e^{i\o \dm }\right)^n-1|\le \sum_{l=0}^n |p_l| |V\left(e^{i\o \dm }\right)-1|^l.
\eaaa
Furthermore, we have that there exists $C_1=C_1(n)>0$ such that
\baaa
\WW(\o,\o_{\dm,\pi,k} ,q,c)^{-n/2}\le C_1(n)\WW(\o,\o_{\dm,\pi,k},q,c)^{-l/2}\brea \forall k=0,1,...,\dm-1, \quad l=1,...,n.
\eaaa
By the choice of the function $\WW$ and by property (iii), it follows   that,
for any $c>0$, there exists $\g_0>0$ and $C=C(n)>0$
    such that, for any $\g>\g_0$,
\baaa
|V\left(e^{i\o \dm }\right)^n-1|
\WW(\o,\o_{\dm,\pi,k},q,n)^{-n/2}d\o\le C(n)\quad\brea  \forall\o\in (-\pi,\pi],\quad \forall k=0,1,...,\dm-1.
\eaaa
This implies that the following analog of the property (b4) holds.
\begin{itemize}
\item[(b4')]
For any $c>0$, there exists $\g_0>0$ and $C=C(n)>0$
    such that, for any $\g>\g_0$,
\baaa
|V\left(e^{i\o \dm }\right)^n-1|
\WW(\o,\o_{\dm,\pi,k},q,c)^{-1/2}d\o\le C(n)\quad  \forall\o\in (-\pi,\pi],\quad \forall k=0,1,...,\dm-1.
\eaaa
\end{itemize}
 Clearly, $\a=\a(\g)\to 1$  and $\mes\, Q(\a)\to 0$ as $\g\to+\infty$.
It follows that, for any $c>0$,
\baa  \|\WW(\o,\o_{\dm,\pi,k},q,c)^{-1/2}\|_{L_2 (Q(\a))}\to 0\quad\brea \hbox{as}\quad \g\to +\infty\quad \forall k=0,1,...,\dm-1.
\label{I1D}\eaa

For $c>0$ and $r\ge 0$, consider an arbitrary  $x\in\X_{m\nu ,\pi}(\qq,c,r)$. We denote  $X\defi \Z x$  and
$\w X\defi \w H_n\,\! X$.

\noxxx{Figure ref{fig2} shows an example of the shape of error curves for approximation of
 the  forward one step shift operator. More precisely, they show
 the shape of $|\w H\ew -e^{i\o n}|$  for the transfer function   (\ref{wH}) with $n=4$,
  and  the shape of the corresponding predicting kernel $\w h$ with some selected  parameters.
}
\def\yo{x_n}
 \par Let $\yo (t)\defi x(t+n)$ and $X_n\defi\Z\yo$. In this case, $X_n(z)=z^n X(z)$. We have that
 \baaa
 \|X_n\ew-\w X\ew\|_{L_1(-\pi,\pi)}=I_1+I_2,
 \eaaa
where \baaa &&I_1=\int_{Q(\a)}| X_n\ew-\w X\ew| d\o,\qquad\breakk
I_2=\int_{Q_+(\a)}|X_n\ew-\w X\ew| d\o. \eaaa

By the assumption, (\ref{hfin}) holds.
 Hence
\baaa &&I_1=\|
\w X\ew- X_n\ew\|_{L_1 (Q(\a))}=
\|(\w H_n\ew-e^{i\o n})X\|_{L_1 (Q(\a))}\nonumber\\
&&\le\sum_{k=0}^{\dm-1}\|(V\left(e^{i\o \dm }\right)^n  -1)
 \WW(\o,\o_{\dm,\pi,k},q,c)^{-1/2}\|_{L_2 (Q(\a))} \|X\ew \WW(\o,\o_{\dm,\pi,k},q,c)^{1/2}\|_{L_2(-\pi,\pi)}.
\label{4s}\eaaa
By the properties of $V$, it follows that $I_1\to 0$ as $\g\to +\infty$
 uniformly over $x\in\X_{m\nu,\pi}(\qq,c,r)$.

Let us estimate $I_2$. We have that
\baaa I_2=\int_{Q_+(\a)} | (e^{i\o n}-\w H_n\,\!\left(e^{i\o  }\right)) X\ew|  d\o\break\le
\int_{Q_+(\a)} |e^{i\o n}(1-V\left(e^{i\o \dm }\right)^n) X\ew|  d\o\\\le
\|1-V\left(e^{i\o \dm }\right)^n\|_{L_2(Q_+(\a))} \|X\ew\|_{L_2(-\pi,\pi)}.\eaaa
\par
Further,
$\Ind_{Q_+(\a)}(\o) |1-V\left(e^{i\o\dm}\right)^n| \to 0$ a.e. as $\g\to
+\infty$. By the properties of $V$, \baaa
\Ind_{Q_+(\a)}(\o)|V\left(e^{i\o \dm }\right)^n-1 | \le 2^n\sum_{l=0}^n|p_l|. \quad \forall \g:\ \g>\g_0.\eaaa   From Lebesgue Dominance Theorem,
it follows that
$\|V\left(e^{i\o \dm }\right)^n-1\|_{L_2(Q_+(\a))}\to 0$ as $\g\to+\infty$. It follows that
$I_1+I_2\to 0$ uniformly over $x\in\X_{m\nu,\b}(q,c,r)$.  Hence
\baa
\sup_{k\in\ZZ}|x(k+n)-\w x(k)|\to 0\quad \hbox{as}\quad \g\to+\infty\quad\brea
\hbox{uniformly over}\quad x\in \X_{m\nu,\b}(q,c,r),
\label{conv}\eaa
 where the process $\w x$ is the output of the linear predictor defined by the kernel $\w h_n=\Z^{-1}\w H_n$ as
 \baa
  \w x(k)\defi \sum^{k}_{p=-\infty}\w h_n(k-p)x(p).\label{predict}
\eaa
Hence the
predicting kernels $\w h_n=\Z^{-1}\w H_n\,\!$ constructed for $\g\to +\infty$
are such as required. This
 completes the proof of Lemma \ref{lemmaX}, i.e. the proof of  Theorem \ref{ThV} for the case where  $\b=\pi$.
\subsubsection*{The case where $\b\in(-\pi,\pi]$ }
We are now in the position to complete the proof of Theorem \ref{ThV} for an arbitrarily selected  $\b\in(-\pi,\pi]$
 \par
 {\em Proof of  Theorem \ref{ThV}}. We have that
 $R_{m\nu,\b}=\{\o_{m\nu,\b,k}\}_{k=0,1,...,m\nu-1}$, where
\baaa
\o_{m\nu,\b,k}=\frac{\b-2\pi k}{m\nu},\quad k=0,1,...,m\nu-1.
\eaaa
Let $\t\defi (\b-\pi)/(m\nu)$. We have that
\baaa
\o_{m\nu,\pi,k}=\o_{m\nu,\b,k}-\t,\quad k=0,1,...,m\nu-1.
\eaaa
Let us select \baaa
\psi(t)\defi e^{i\t t},\qquad x_\b(t)\defi \psi(t)x(t),\qquad t\in\ZZ.
\eaaa
Let $x\in \X_{m\nu,\b}(q,c,r)$ for some $\qq>0$ and and integer $\nu>0$. Let $X_\b=\Z x_\b$ and $X=\Z x$.
In this case,
\baaa
X_\b\ew=\sum_{k\in\ZZ}e^{-i\o k}x_\b(k)=\sum_{k\in\ZZ}e^{-i\o k}\psi(k)x(k)\\
=\sum_{k\in\ZZ}e^{-i\o k +i\t k}x(k)=X\left(e^{i\o-i\t}\right).
\eaaa
We have that
\baaa
X_\b\left(e^{i\o_{m\nu,\pi,k} }\right)=X\left(e^{i\o_{m\nu,\pi,k}-i\t}\right)=X\left(e^{i\o_{m\nu,\b,k}}\right)=0,\quad k=0,1,...,m\nu-1.
\eaaa
Hence $x_\b\in \X_{m\nu,\pi}(q,c,r)$. As we have established above,  $x_\b$  is predictable in the sense of Definition \ref{defP}. Let $\w h_n$ be the corresponding
predicting kernel which existence  is required by Definition \ref{defP} for the case where $\psi(t)\equiv 1$ such that the approximation $\w x_\b(t)$ of $x_\b(t+n)$
 is given by
  \baaa
\w x_\b(t)=\sum_{k=-\infty}^t \w h_n(t-k)y_\b(k).
\eaaa
These kernels
 were
 constructed in the proof of Theorem \ref{ThV}  above for the special case   $\b=\pi$.
It follows that  \baaa
\w x(t)=\psi(t)^{-1}\sum_{k=-\infty}^t \w h_n(t-k)\psi(t) x_(k).
\eaaa
 This completes the proof of Theorem \ref{ThV}.
$\Box$
\par
{\em Proof of Theorem \ref{ThR}.}
It suffices to show that the error for recovery a singe term $x(n)$ for a given integer $n>0$  from the sequence $\{x(mk)\}_{k\in \ZZ,\, k\le 0}$ can be made arbitrarily small is a well-posed problem; the proof for a finite set of values  to recover is similar.

The case where  $n<0$ and $\{x(mk)\}_{k\in \ZZ,\, k\ge 0}$ are observable
can be considered similarly.

We assume that $N>n$.

Let us consider an input sequence $x\in\ell_2$
such that \baa
x=\ww x+\eta,\label{xeta}
\eaa
where    $\eta(k)= \xi(k)\Ind_{\{|k|\le N\}}-\ww x(k)\Ind_{\{|k|>N\}}$.
In this case, (\ref{xeta}) gives that
$x(k)=(\ww x(k)+\xi(k))\Ind_{\{|k|\le N\}}$.

Let  $X=\Z x$, and let ${\cal N}\defi \Z\eta$, and let $\s=\|{\cal N}\ew\|_{L_2(-\pi,\pi)}$; this parameter  represents the
intensity of the noise.

Let us assume first that $\b=\pi$.

Let  an  arbitrarily small $\e>0$ be given.
Assume that  the parameters $(\g,\w r)$ of $\w H_{n}$ in (\ref{wH}) with $\nu=\nu_d$
are selected such that \baa
|\oo x(n)-\oo x(0)|\le \e/2 \quad \forall  \oo x\in \X_{m\nu,\b}(\qq,c,r),
\label{yy}\eaa
for  $\oo x=\Z^{-1}(\w H_n\oo X)$ and $\oo X=\Z\oo x$.
The kernel $\w h_n$ produces an estimate $\oo x(0)$  of
based on observations of $\{\oo x(km)\}_{k\le 0}$.

 Let  $\ww x\defi\Z^{-1}(\w H_n\ww X)$, where $\ww X=\Z \ww x$. By (\ref{yy}), we have that
\baaa  |\ww x(0)-\ww x(n)|\le\e/2.
\label{eps}\eaaa

 Let  $\w x\defi\Z^{-1}(\w H X)$, where $X=\Z x$.  We have that
$\w x=\ww x+\Z^{-1}(\w H {\cal N})$ and
 \baaa
  |\w x (0)-x(n)|\le   |\ww x (0)-x(n)|+|\Z^{-1}(\w H {\cal N})(0)|\le \e/2+ E_\eta,
  \eaaa
where
\baaa
E_{\eta}\defi \frac{1}{2\pi}\|(\w H\ew-e^{i\o n}){\cal N}\ew|\|_{L_1(-\pi,\pi)}\le \s (\kappa+1),
\eaaa
and where
\baaa
\kappa\defi \sup_{\o\in[-\pi,\pi]}|\w H_n\ew|.\eaaa
\par
We have  that  $\s\to 0$ as $N\to +\infty$ and $\|\xi\|_{\ell_2(-N,N)}\to 0$. If $N$ is large enough and  $\|\xi\|_{\ell_2(-N,N)}$ is small enough  such that $\s (\kappa+1)<\e/2$, then
$|\w x(0)-   x(n) |\le\e$.
This completes the proof of Theorem \ref{ThR} for the case where $\b=\pi$.

The prove for the case where  $\b\neq \pi$ follows from the case where $\b=\pi$, since
the processes $x(t),\oo x(t),\ww x(t),\w x(t),\ww x(t),$  presented in the proof  with $\b=\pi$ will be simply
converted to  processes $e^{-i\t t}x(t),e^{-i\t t}\oo x(t),e^{-i\t t}\w x(t),e^{-i\t t}\ww x(t)$, for $\t=(\b-\pi)/(m\nu)$, similarly the proof of Theorem \ref{ThV}.
 $\Box$
\begin{remark}{\rm
By the properties of $\w H_n$, we have that $\kappa\to +\infty$  as $\g\to +\infty$.
 This implies that error (\ref{eps}) will be increasing if $\w\e\to 0$ for any given $\s >0$.
This means that, in practice, the predictor  should not  target too small a size of the
error, since in it impossible to ensure that $\s=0$ due inevitable data truncation.
 }\end{remark}

\par
{\em Proof of Lemma \ref{lemmaSS}}.
 Let $y\in \Y_{m,\nu,\b,s}(\qq,c,r)$   for some $m,\nu,\b,\d,s,r$. By the definitions, there exists   $x\in\X_{m\nu,\b}(\qq,c,r)$ such that  $y=\SS_{m,s} x$.  It suffices to prove that $y\in \X_{m\nu,\b}(\qq,c,r_1)$ for some $r_1=r_1(m,\nu,\b,\d,s,r)$.

Let us consider the case where $s=0$. In this case, $Y=\Z y$ can be represented as
\baaa
Y\ew =\frac{1}{m}\sum_{k=1}^m X\left(e^{i\o+i\o_{m,0,k}}\right). \eaaa
(See e.g. \cite{Oppen} ,  pp. 167-168).   To prove that $y\in \X_{m\nu,\b}(\qq,c,r_1)$ for some $r_1>0$, it suffices to show that
\baa
e^{i(\o_{m\nu,\b,l}+\o_{m,0,k})}\in  \{e^{i \o_{m\nu,\b,d}}\}_{d=0,1,..,m\nu-1},\quad k,l\in\ZZ.
\label{ein}\eaa
We have that
\baaa
\o_{m\nu,\b,l}=\frac{2\pi l-\b}{ m\nu},\quad \o_{m,0,k}=\frac{2\pi k}{ m}.
\eaaa
Hence for all  $k,l\in\ZZ$ we have that
\baaa
e^{i(\o_{m\nu,\b,l}+\o_{m,0,k})}=\exp\left( i\frac{2\pi l-\b}{ m\nu}+i\frac{2\pi k}{ m}  \right)
=\exp\left( i\frac{2\pi l-\b+2\pi km }{ m\nu}\right).
\eaaa
This implies that (\ref{ein}) holds.

Consider now a general case of $s\in\ZZ$.

Consider operator $S:\ell_2\to\ell_2$ such that
$(Sx)(t)=x(t+s)$.  Let $x_s=Sx$ and $y_s=Sy$.
For $y=\SS_{m,s} x$, we have that $y_s=Sy=\SS_{m,0} x_s$.  Clearly,  $x_s\in \X_{m\nu,\b}(\qq,c)$.
As was proved above,  $y_s\in \X_{m\nu,\b}(\qq,c,r_1)$ for some $r_1>0$. Hence $y\in \X_{m\nu,\b}(\qq,c,r_1)$.
This completes the proof of Lemma \ref{lemmaSS}.$\Box$

\par
The proof of Corollary \ref{corrM}. For $\X=\Y_{m,\nu,\b,s}(\qq,c,r)$, the proof follows immediately from the proof of Lemma \ref{lemmaSS}. The proof for
$\X=\Y_{m,\nu,\b,s}(\qq,c,r)$ follows immediately from the observation that there is a natural bijection
between $\Y_{m}$ and $\w\Y_{m}$; one may simply represent $\w y\in\w\Y$ as an element of $\Y$  and establish predictability  by application of the predictor derived above for sequences from $\Y_m$. $\Box$

\par
{\em Proof of Corollary \ref{corrSUB}}.
It suffices to
prove that if $y\in \Y_{m}$  such that $y(k)=0$ for
$k\le s$, then $y(k)=0$ for $k>q$ for all $q\in\ZZ$.
By the predictability established by Theorem \ref{corrM}, for a process $y\in\Y_m$ such that
$y(k)=0$ for $k\le s$, it follows that $y(k+1)=0$;  we obtain this by letting $\g\to +\infty$
in (\ref{wH}).  Similarly, we obtain that $y(k+2)=0$.
Repeating this procedure,
we obtain that $y(k)=0$ for all $k\in\ZZ$. $\Box$
\par

{\em Corollary \ref{corr2}} follows immediately from Corollary \ref{corrSUB}. $\Box$

{\em Lemma \ref{lemmaSD}}. Let $\w y\in\w\Y_{m,s}$ and let $\w Y=\Z \w y$.
By the definitions, $\w y=\SSS_{m,s}x$ for some $x\in \X_{m\nu,\b}(\qq,c,r_1)$ and $\nu,\b,r_1$.  Let $s=0$, and let  $Y(z)\defi \w Y(z^{m})$.
By Lemma \ref{lemmaSS}, $y\in \X_{m\nu,\b}(\qq,c,r_1)$. At the same time,  $y=\SupS_{m,0} \w y$.  Then the  proof follows.

 then,  follows fro $\Box$

 The proof of Theorem \ref{ThM2}, Corollary \ref{corrSUB2}, and Corollary \ref{corr22}
 repeats the proofs above with minor modifications, and will be omitted here.

\par
{\em Proof for Example \ref{ex1}}.  It suffices to show that
  $\o_{\mu(m,d_1),k}\neq \o_{\mu(m,d_2),l}$ if $d_1\neq d_2$ for all $k,l\in\ZZ$ , where
  $\o_{n,k}=\frac{2\pi k-\pi}{ n}$ are the roots of the equation $e^{inw}=-1$.
 Suppose that
$\o_{\mu(m,d_1),k}= \o_{\mu(m,d_2),l}$ for some $k,l\in\ZZ$ for $d_2=d_1 +r$, for some
 $d_1,d_2\in\{-m+1,...,m-1\}$ such that $a\defi d_2-d_1> 0$. In this case, the definitions imply that
\baaa
 \frac{2 k-1}{ 2^{d_1}}=\frac{2 l-1}{2^{d_2} }.
 \eaaa
This means that  $ 2^a(2 k-1)=2 l-1$ which implies that the number $2^a(2 k-1)$ is odd. This is impossible since we had assumed that  $a>0$. Hence
 the sets $R_{m\nu_d,\b}$ are mutually disjoint for different
 $d$.  This completes the proof for Example \ref{ex1}. $\Box$

\def\TTT{{\cal I}}
We will need an extended version of Definition \ref{defR}.

\begin{definition}\label{defP3}
Let $P$ be a finite subset of $\ZZ$ with $N>0$ elements,
 Let $\X$ be a class of ordered sets $\{x_p\}_{p\in P}\subset (\ell_2)^N$.   Let a set $\TTT\subset \ZZ$ be given, and let $\TTT(M)= \TTT\setminus \ZZ_{[-M,M]}$.
We say that the class $\oo\Y$ is  uniformly recoverable from observations of $x_0$ on  $\ww\TTT$  if, for any integer $M>0$ and any $\e>0$, there exists a set $\{\wh_{p,t}(\cdot)\}_{p\in P,t\in\ZZ_{[-M,M]}}\subset \w\K$ such that \baaa \max_{p\in P,\ t\in\ZZ_{[-M,M]}} |x_p(t)- \w x_p(t)|\le \e\quad
\forall x\in\X, \label{predU2}\eaaa where  \baaa
\w x_p(t)=
\sum_{s\in\TTT(M)}\wh_{p,t}(t-s)x_0(s).\eaaa
\end{definition}

Let us introduce  mappings $\M_d:\ell_2\to \ell_2$ for $d\in\ZZ_{[-m+1,m-1]}=\{-m+1,...,m-1\}$
such that, for  $x\in\ell_2$,  the sequence $x_d=\M_d x$ is defined by insertion of $|d|$ new members equal to
$x(0)$    as the following:
\baaa
&&\hphantom{}\hbox{(i)}\,\,\,\, x_0=x; \hphantom{xxxxx}  \hphantom{x_d(k)=x(0),\quad k= 0,1,...,d,\quad k<0}\,\,\nonumber \\
&&\hphantom{}\hbox{(ii)\,\, for} \quad d>0:\quad \breakk \hphantom{xxxxx}  x_d(k)=x(k),\quad k<0, \nonumber\\
&&\hphantom{xxxxx} x_d(k)=x(0),\quad k= 0,1,...,d, \nonumber\\
&&\hphantom{xxxxx} x_d(k)=x(k-d),\quad k\ge d+1;\nonumber\\
&&\hphantom{}\hbox{(iii) \,\, for} \quad d<0:\quad \breakk\hphantom{xxxxx}  x_d(k)=x(k),\quad k>0, \nonumber\\
&&\hphantom{xxxxx} x_d(k)=x(0),\quad k= d,d+1,...,0, \nonumber\\
&&\hphantom{xxxxx} x_d(k)=x(k-d),\quad k\le d-1.\hspace{-1cm}\label{xd2}
\eaaa

\par

{\em Proof of Theorem \ref{ThDeg}}.
Let  $\nu_d>0$  be some integers $d\in\ZZ_{[-m+1,m-1]}=\{-m+1,...,m-1\}$. Let $\ww\Y_d$ be the set of  $\xi_d=\SS_{m,0} x_d$, where  $x_d=\M_d x$, for all
$x\in \PP_{m,\nu,\b}(\qq,c,r)$.

Let $\ww\Y(\d)$ be the set of all ordered sets $\{\xi_d\}_{d=-m+1}^{m-1}$ such that $\xi_d\in \ww\Y_d(\d)$.

By Lemma \ref{lemmaX}, the sets $\ww\Y_d(\d)$  are  uniformly predictable in the sense of Definition
 \ref{defP}. It follows that,  for any $s\in\ZZ^+$ and $d\ge 0$, the sets $\ww\Y_d(\d)$  are  uniformly recoverable  in the sense of Definition
 \ref{defR} from observations   of $\{\xi_d(k)\}$ on $\{k\in \ZZ:\ k\le -s\}$.  Clearly, time direction
 can be reversed and therefore it can be concluded that, for $d<0$ and any $s\in\ZZ$, the sets $\ww\Y_d(\d)$  are  uniformly recoverable  in the sense of Definition   \ref{defR} from observations of $\{\xi_d(k)\}$ on
$\{k\in \ZZ:\     |k|\ge s\}$.  By the structure of $\xi_d=\SS_{m,0} x_d$,
 it follows that it sufficient to use observations of $\xi_d$ on $\{k\in \ZZ:\  k/m\in\ZZ,\   k\le -s\}$ only for $d\ge 0$ and, respectively,  observations of $\xi_d$ on $\{k\in \ZZ:\  k/m\in\ZZ,\   k\ge s\}$ only for $d< 0$.

  By the definitions, we have for  $d>0$ that  $\xi_d(k)=\xi_0(k)$ for $k\le 0$.
 For  $d<0$,  we have that $\xi_d(k)=\xi_0(k)$ for $k\ge 0$.  Hence, for any $s>0$,  the set $\ww\Y(\d)$  is  uniformly recoverable  in the sense of Definition   \ref{defP3} (with $\psi_p\equiv 1$) from observations of $\xi_0$ on the set  $\TT_0(s)=\{k\in \ZZ:\ k/m\in\ZZ,\ |k|\ge s\}$.

By the definitions again, we have that $\xi_0(km)=x(km)$ for all $k\in\ZZ$. For  $d>0$,  we have that $\xi_d(km-d)=x(km-d)$ for $k\ge 1$ and $\xi_d(km)=x(km)$ for $k\le 0$.
  For  $d<0$,  we have that $\xi_d(km-d)=x(km-d)$ for $k\le -1$ and $\xi_d(km)=x(km)$ for $k\ge 0$.
 Hence $x$  is  uniformly recoverable  in the sense of Definition   \ref{defR} from
 from observations of $x$ on $\TT_0(s)$. This  completes the proof of Theorem \ref{ThDeg}. $\Box$

{\em Proof of Theorem \ref{ThDense}}.  In the proof below, we consider $d\in\ZZ_{[-m+1,m-1]}=\{-m+1,...,m-1\}$.

 Let $x\in\ell_2$ be arbitrarily selected and let
$x_d\defi \M_d x$.

We assume that  all $m\nu_d$ are different for different
 $d$.
\par
Let $Y_d\defi X_0-X_d$ and $\xi_d=\Z^{-1}Y_d$, where $ X_d=\Z  x_d$.
Since $x_d(k)=x_0(k)$ for $k\le 0$ and $d>0$, and $x_d(k)=x_0(k)$ for $k\ge  0$ and $d<0$, it follows that $\xi_d(k)=0$ for $k\le 0$ and $d>0$, and $\xi_d(k)=0$ for $k\ge 0$ and $d<0$. In addition, $Y_0\equiv 0$ and $\xi_0\equiv 0$.

Further, let $D=\{k\in\ZZ:\ |k|\le m-1,\ k\neq 0\}$, and let
\baaa A(\o)=\prod_{d=-m+1}^{m-1}\prod_{k=0}^{m\nu_d-1}\varrho(\o,\o_{m\nu_d,\b,k},q,c)^{-1},\quad
\a_d(\o)=\prod_{k=0}^{m\nu_d-1}[1-\varrho(\o,\o_{m\nu_d,\b,k},q,c)^{-1}].
\eaaa

We have that,  for any $(L,\w r,c)$,  for large enough $\g$,
\baa (a_d(\o)-1)\varrho(\o,\o_{m\nu_d,\b,k},q,c)\equiv 1\quad k=0,1,...,\nu_d-1.
\label{arho}\eaa
We assume that $\g,L,\w r,c$ are selected  such that (\ref{arho}) holds.

Let \baaa
\w X_0\ew\defi  X_0\ew A(\o)\brea +
\sum_{p\in D} Y_p\ew a_p(\o).
\eaaa We have that $\w x_0\in \X_{md_0,\b}(q,c,r)$, where $\w x_0=\Z^{-1}\w X_{0}$.

Furthermore,  by the definitions,
\baaa
\w X_d\ew=X_d\ew+\w X_{0}\ew-X_0\ew=X_0\ew A(\o)+W_d\ew,
\eaaa
where
\baaa
W_d\ew=X_d\ew+
\sum_{p\in D} [X_0\ew-X_p\ew] a_d(\o) -X_0\ew.
 \eaaa
For large enough $L>0$, the function $\WW$ takes the value one on the most part of $\TT$, with some isolated peaks.
Because of this,  we have that
\baaa a_p\ew\varrho(\o,\o_{m\nu_d,\b,k},q,c)=0\quad\hbox{a.e.},\quad p\neq d,\quad k=0,1,...,\nu_d-1.
\eaaa
By the choice of $A\ew$, it follows  for all $d$ that
\baa \max_{k=0,1,...,m\nu_d-1}\int_{-\pi}^\pi |X_0\ew A(\o)|^2 \WW(\o,\o_{m\nu_d,\b,k},q,c)^2d\o\le r,
\label{XA}\eaa\ where $X=\Z x$.

Let us show that,  for large enough $\g$, \baa \max_{k=0,1,...,m\nu_d-1}\int_{-\pi}^\pi |W_d\ew|^2 \WW(\o,\o_{m\nu_d,\b,k},q,c)^2d\o\le r,
\label{W}\eaa where $X=\Z x$.
We have that
\baaa
W_d\ew=X_d\ew-X_0\ew+
\sum_{p\in D} [X_0\ew-X_p\ew] a_p(\o)\\
= Y_d\ew(a_d(\o)-1)+\sum_{p\in D,\ p\neq d} Y_p\ew a_p(\o)  .
 \eaaa

 By (\ref{arho}), it follows that (\ref{XA}) and (\ref{W}) holds. Hence
\baa \max_{k=0,1,...,m\nu_d-1}\int_{-\pi}^\pi |X_d\ew|^2 \WW(\o,\o_{m\nu_d,\b,k},q,c)^2d\o\le r
\label{1}\eaa
 and $x_d\in \X_{m\nu_d,\b}(q,c,r) $ for all $d$.
%\end{document}

In addition, it was shown above that $\w x_d(k)=\w x_0(k)$, for $k\le 0$ and
 $d>0$, and  $\w x_d(k)=\w x_0(k)$, for $k\ge 0$ and
 $d<0$.
 Let $\w x$
 be defined by (\ref{xPm}), where $\xi_d= \w x_d$.
 By the definitions, it follows that
   $\w x\in\PP_{m,\nu,\b}(\qq,c,r)$.

Further, assume that $q>1$ is fixed and $c\to 0+$. Clearly, \baaa
 &&\|a_p\|_{L_2(-\pi,\pi)}\to 0,\quad
 \|A\|_{L_2(-\pi,\pi)}\to 1\quad \hbox{as}\quad c\to 0+ .
\label{aA}\eaaa
 It follows that  \baaa
 &&\|\w X_0\ew-X_0\ew\|_{L_2(-\pi,\pi)}\to 0,\\
 &&\|\w X_d\ew-X_d\ew\|_{L_2(-\pi,\pi)}\to 0\quad  \hbox{as}\quad c\to 0+ .
\label{aAA}\eaaa
It follows that \baa
 \|\w x_d-x_d\|_{\ell_2}\to 0\quad \hbox{as}\quad c\to 0.
\label{xdd1}\eaa
By the definitions, it follows that
\baaa
\w x(k)-x(k)=\w x_d(k+d)-x_d(k+d),\quad \brea k\ge 0,\quad d\ge 0,\quad (k+d)/m\in\ZZ,
\eaaa
and
\baaa
\w x(k)-x(k)=\w x_d(k+d)-x_d(k+d),\quad \brea k<0,\quad d<0,\quad (k+d)/m\in\ZZ.
\eaaa
Then (\ref{xdd}) follows from (\ref{xdd1}). This completes the proof of Theorem \ref{ThDense}. $\Box$
%%%%%%%%%%%%

\par
{\em Proof of Theorem \ref{ThRR}} follows from Theorem \ref{ThR} applied to the prediction of the corresponding subsequences. $\Box$

\subsection*{Some properties of predicting kernels}
\begin{proposition}\label{proph}
\begin{enumerate}
\item The predicting kernels $\w h_n$ are real valued,
\item The predicting kernels $\w h_n$  are quite sparse:
\baa
 \w h_n(k)=0\quad  \hbox{if either}\quad (k+n)/(m\nu)\notin \ZZ\quad \hbox{or}\quad  k<mn-n.
 \label{hm}\eaa
 \end{enumerate}
\end{proposition}
\par
{\em Proof.} To prove (i), it suffices to observe that    $\w H_n\,\!\left(\oo z\right)=\overline{\w H_n\,\!\left(z\right)}$.
Let us prove (ii).
By the choice of $V$, it follows that $|V(z)|\to 0$ as $|z|\to +\infty$. Hence  $v(0)=0$ for $v=\Z^{-1}V$ and \baaa
&&V(z^\dm)\breakk =z^{-\dm}v(1)+z^{-2\dm}v(2)+z^{-3\dm}v(3)+.. .\eaaa
 Clearly, we have that
\baaa
V(z^\dm)^n=z^{-n\dm}w(1)+z^{-(n+1)\dm}w(2)+z^{-(n+2)\dm}w(3)+...,\eaaa
where $w=\Z^{-1}(W)$ for $W(z)=V(z)^n$.
Hence \baaa
&&\w H_n(z^\dm)=z^nV(z^\dm)^n\breakk=z^{n-n\dm}w(1)+z^{n-(n+1)\dm}w(2)+z^{n-(n+2)\dm}w(3)+...\\
&&= z^{n-n\dm}\w h(n\dm-n)+z^{n-(n+1)\dm}\w h((n+1)\dm-n)\breakk+z^{n-(n+2)\dm}\w h((n+2)\dm-n)+...
\eaaa
Hence (\ref{hm}) holds. This completes the proof of Proposition \ref{proph}. $\Box$

\section{On numerical implementation}
%\subsubsection*{On the growth of $\|\w h_n\|_{\ell_\infty}$}

Our numerical experiments show  that $\|\w h_n\|_{\ell_\infty}$ is growing very fast as $\g$ is increasing.
For the case of relatively small $\g\sim 2$, some experiments were done in \cite{DH16}. These experiments
demonstrated that it is possible to achieve small but noticeable improvement of the prediction accuracy for autoregressions in stochastic setting.
In the present paper, we attempted to improve approximation
using larger $\g$. The main challenge here is   that  $\|\w h_n\|_{\ell_\infty}$  is growing very fast as
 is increasing very fast as $\g$ is increasing.

\begin{figure}[ht]
\centerline{\psfig{figure=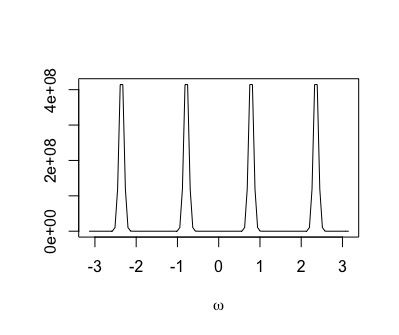,width=9cm,height=6.0cm}}
\caption[]{\sm The distance $|\w H_4\ew-e^{2 i\o}|$ for $\g=3$ on $[-\pi,\pi]$.
 }\label{fig1}
  \centerline{\psfig{figure=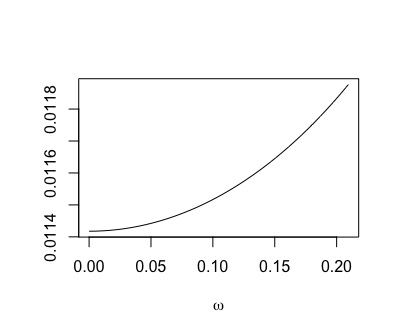,width=9cm,height=6.0cm}}
\caption[]{\sm The distance $|\w H_4\ew-e^{2 i\o}|$ for $\g=3$ on a interior interval inside
$(-\pi/4,\pi/4)$.
 }\label{fig2}
  \centerline{\psfig{figure=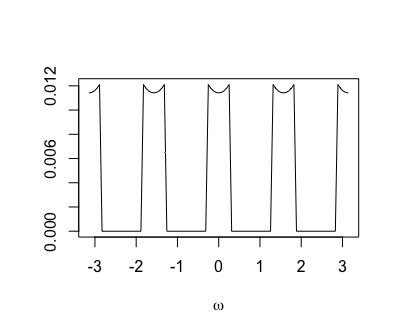,width=9cm,height=6.0cm}}
\caption[]{\sm The distance $|\ww H_4\ew-e^{2 i\o}|$ for $\g=3$ on $[-\pi,\pi]$. }\label{fig3}

 \vspace{0cm}\end{figure}

 For example, we calculated  for $n=2$, $r=1.2$, the kernel $\w h_n$  for different choices of $\gamma$ using standard
integration in {\em R}. We obtained that $\|\w h_n\|_{\ell_\infty}=239,190$ for $\gamma=3$, that
$\|\w h_n\|_{\ell_\infty}= 2,985,964\cdot 10^{29}$
 for  $\gamma=6$, and that $\|\w h_n\|_{\ell_\infty}= 2,475,003\cdot 10^{122}$.
 for  $\gamma=10$. \index{T=251}  Figures \ref{fig1} and \ref{fig2} shows
 the distance $|\w H_2\ew-e^{2 i\o}|$ from the transfer function $e^{2i\o}$ of the ideal two-step predictor for $\g=3$.

This makes calculations with large $\g$

 The values of $\w h_n$ of this magnitude  are being coded  in a standard computer programm with
  a large error; this generates a large error for the prediction.
 Potentially, application of more precise computational
 technique with an adequate number of digits for representation for large numbers.

The theorems presented above focus on processes without spectrum gaps
of positive measure on $\T$. However, for numerical examples, we considered
 more special processes with periodic spectrum gaps of a positive measure  $\T$
 defined as the follwing.

For $\d>0$, $\b\in (-\pi,\pi]$,   $\w\o\in(-\pi,\pi]$, and $n\in\ZZ_1^+$,  let \baaa
  &&J(\w\o,\d)=\{\o\in (-\pi,\pi]: \quad |e^{i\o}-e^{i \w\o}|\le \d\},
  \breakk\quad J_{n,\b}(\d)=\cup_{k=0}^nJ\left(\o_{n,\b,k},\d\right).
\label{J}\eaaa

Let $\V_{n,\b}(\d,r)$ be the set of all   $x\in B_r(\ell_2)$ such that
$X\ew=0$ for a.e. $\o\in J_{n,\b}(\d)$, where $X=\Z x$. These processes have periodic spectrum gaps
of positive measure on $\T$.

Clearly, $\V_{n,\b}(\d,r)\subset \X_{n,\b}(q,c,r_1)$ for some $r_1=r_1(r,q,c,\d,n)$, and there exists
$C=(r,q,c,\d,n) >0$ such that \baaa
\int_{-\pi}^\pi |X\ew|^2 d\o\le\max_{k=0,1,...,n-1}\int_{-\pi}^\pi |X\ew|^2 \WW(\o,\o_{n,\b,k},q,c)^2d\o\le C_1\int_{-\pi}^\pi |X\ew|^2 d\o
\eaaa
for all $x\in \V_{n,\b}(\d,r)$ and $Z=\Z x$.

So far, we managed to make some experiments allowing to establish what is  a sufficient size of $\g$
for  prediction of processes from $x\in\V_{2,\pi}(\d)$, i.e. with periodic spectrum gaps on $\T$ having a  positive measure given that we are able to deal with large values of kernels $h_n$.
For this, we done experiments that allowing to recreate  this scenario without actually using large values of $h_n$.  Let us describe this experiments.

  Let \baaa
 I(\o)\defi \Ind_{J_{n,\b}(\d)}(\o),\quad \ww H_n\ew \defi \w H_n\ew I\ew,\quad \ww h_n\defi\Z^{-1}\ww H_n.
 \eaaa
 The idea is to replaced causal predicting kernels  $\w h_n$ by more regular non-causal kernels $\w h_n$
 that,  in the theory,  would lead to the same results  $x\in\V_{2,\pi}(\d)$, if were able to complete calculations with large $\w h_n$ and large $\g$  with sufficient precision.
Let us explain why the results would be the same.  For  $x\in\V_{2,\pi}(\d)$ and
 \baaa
 \w x(t)=\sum_{k\le t} \w h_n(t-k) x(k),\quad \w X=\Z \w x,
 \eaaa
  we have that
 \baaa
 &&\w x(t)=\frac{1}{2\pi}\int_{-\pi}^\pi e^{i \o t}\w X\ew  d\o=\frac{1}{2\pi}\int_{-\pi}^\pi e^{i \o t}\w H_n\ew X\ew d\o
 \\&&=\frac{1}{2\pi}\int_{-\pi}^\pi e^{i \o t}\w H_n\ew X\ew I\ew d\o=\frac{1}{2\pi}\int_{-\pi}^\pi e^{i \o t}\ww H_n\ew X\ew d\o\\
 &&=\sum_{k\in\ZZ} \ww h_n(t-k) x(k).
 \eaaa

Figure \ref{fig3}  shows
 the distance $|\ww H_2\ew-e^{2 i\o}|$  from the transfer function $e^{2i\o}$ of the ideal two-step predictor for $\g=3$.
 Since  $\ww H_2\ew=\w H_2\ew$ for
$\o\in(-\pi,\pi]\setminus J_{4,\pi}(\d)$ for $\d=0.5$, it follows that Figure \ref{fig2}  shows
 the distance $|\w H_2\ew-e^{2 i\o}|$ on a interior interval inside
$(-\pi/4,\pi/4)$.

With this relaxation of the conditions of the experiments, we found that, for large $\g$,
 the value  $\w x(t)$ approximates $x(t+n)$ quite effectively.

\begin{figure}[ht]
\centerline{\psfig{figure=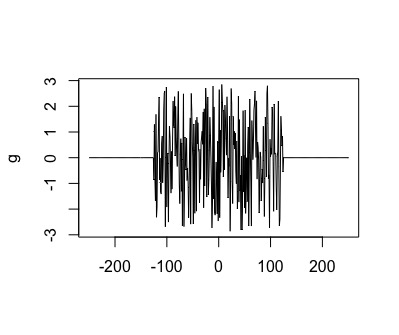,width=9cm,height=6.0cm}}
\caption[]{\sm
Process $g$: a truncated  path a Gaussian process.
 }\label{fig4}
  \centerline{\psfig{figure=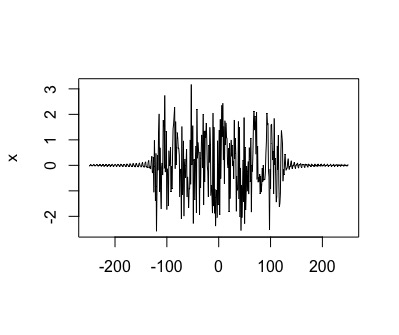,width=9cm,height=6.0cm}}
\caption[]{\sm
 $x=\Z^{-1}(\Ind_{J_{4,\b}(\d)}\ew G\ew)$, $G=\Z g$.
 }\label{fig5}
 \vspace{0cm}\end{figure}

Let us illustrate this on the case of  two-step prediction, i.e. prediction of $x(2)$ given truncated observations $\{x(t)\}_{t/2\in\ZZ, -T\le t\le 0\ }$ for $x\in\V_{4,\pi}(0.5)$.
In our notations, this means that $n=2$ and  $\w x(0)$ is the estimate of $x(2)$.
To create a process $x$,  we created first  a path $\{g(t)\}_{t=-T}^{T}$ generated as a path of a Gaussian stationary process, and we considered $x=\Z^{-1}(\Ind_{J_{4,\pi}(\d)}\ew G\ew)$ for $\d=0.5$, where
 $G=\Z g$; by the definitions, it follows that $x\in\V_{4,\pi}(0.5)$.
 Figure \ref{fig4}  shows the path of $g$, and Figure \ref{fig3}  shows the path of $x$.

We calculated the relative errors
\baaa
E=\frac{|\w x(0)-x(2)|}{\sqrt{\frac{1}{T+1}\sum_{t=-T}^0 x(t)^2}}.
\eaaa
 In particular, for $T=250$ and $\w r=1.2$, we have that
 \begin{enumerate}
 \item
 $E= -0.57$ for $\gamma=3$;
 \item
$E= -0.036$ for  $\gamma=10$;
 \item
 $E=-0.027$ for  $\gamma=20$.
\end{enumerate}

\xxxonly{}
%\end{document}

\end{document}